\documentstyle[12pt]{article}
\renewcommand {\c}  {\'{c}}
\newcommand {\cc} {\v{c}}
\newcommand   {\s}  {\v{s}}
\begin{document}
\pagestyle{empty}
\vspace* {13mm}
\baselineskip=12pt

% ****************************TITLE********************************************
%
\begin{flushright}
{\bf RBI-TH-2/94}\\
{\bf PMF-TH-1/94}\\
{\bf January 1994}
\end{flushright}
\vskip2cm
\begin{center}

 {\bf MAPPINGS OF MULTIMODE BOSE ALGEBRA PRESERVING NUMBER OPERATORS}
  \\[10mm]

M. Dore\s i\c$^1$ , S.Meljanac$^1$  and M.Milekovi\c$^{2,+}$\\[7mm]
{\it $^1$ Rudjer Bo\s kovi\c \ Institute, Bijeni\cc ka c.54, 41001 Zagreb,
Croatia\\[5mm]
 $^2$ Prirodoslovno-Matemati\cc ki Fakultet, Department of Theoretical
Physics\\Bijeni\cc ka c.32, 41000 Zagreb, Croatia\\[5mm]
$^+$ e-mail: marijan@phy.hr}\\
\vskip 3cm
\end{center}
%
%
%**************************ABSTRACT*****************************************

\begin{center}
{\bf Abstract}
\end{center}
We define a class of deformed multimode oscillator algebras which possess
number operators and can be mapped to multimode Bose algebra.
We construct and discuss the states in the Fock space and the
corresponding number operators.
%
%*********************END OF THE ABSTRACT**********************************%
%
%
\newpage
\baselineskip=24pt
\setcounter{page}{1}
\pagestyle{plain}
\def\leer{\vspace{5mm}}
\setcounter{equation}{0}%
%
%
%%**********************BEGIN OF THE PAPER********************************

Recently, much interest has ben devoted to the study of quantum groups
\cite{dj} and to generalizations (deformations) of oscillator algebras \cite
{pw}\cite{bm}\cite{h}\cite{okkcs}\cite{ac}. Any type of single-mode
deformed oscillator  can be related (mapped) to a single-mode
Bose oscillator . Deformations of multimode oscillators have also
been studied \cite{j}. For Pusz-Woronowitz oscillators covariant under the
$SU_{q}(n) $($SU_{q}(n|m) $) quantum  algebra (superalgebra) \cite{pw}\cite{j}
and anyonic-type algebras \cite{lsbdm} there exist mappings to
multimode Bose algebra. However, there are deformed generalized
quon algebras \cite{g} for which such mappings do not exist. It is essential
that all these algebras are associative and the R-matrix approach to them
has been pursued \cite{fz}. Multimode deformed oscillator algebras are
important in possible physical applications, for example in q-deformed
field theory \cite{lsbdm} and generalized statistics \cite{mp}.\\
Our aim is to define and analyze a general class of deformed multimode
oscillator algebras which possess well-defined number operators
for each type of oscillator, and
 can be mapped to a multimode Bose algebra.
These mappings can be viewed as generalized Jordan-Wigner/Klein
transformations. We discuss the corresponding algebras. For special
values of parameters, one recovers all known such algebras \cite{pw}\cite{bm}
\cite{okkcs}\cite{ac}\cite{j}\cite{lsbdm}\cite{fz}. Finally, we construct and
discuss the
 states in the Fock space and the corresponding number operators for deformed
oscillators.

Let us define the annihilation and creation operators $b_{i}$, $b_{i}^{+}$, $
(i \in S)$
satisfying the Bose algebra

\begin{equation}
 \begin{array}{c}
 [b_{i},b_{j}^{+}]=\delta_{ij},\quad \forall i,j \in S\\[4mm]
[b_{i},b_{j}]=0
\end{array}
\end{equation}
where $S={1,2,...n}$ or S is a set of sites on a lattice. The number operators
$N_{i}$ satisfy
\begin{equation}
\begin{array}{c}
[N_{i},b_{j}]=-b_{i} \delta_{ij},\quad \forall i,j \in S\\[4mm]
[N_{i},b_{j}^{+}]=b_{i}^{+} \delta_{ij}\\[4mm]
N_{i}=b_{i}^{+}b_{i}, \quad \forall i,j \in S.
\end{array}
\end{equation}

Now we define generalized Jordan-Wigner transformations of the above
Bose algebra, equations (1) and (2), as

\begin{equation}
a_{i}=b_{i} e^{\sum_{j}c_{ij}N_{j}}\sqrt{\frac {\varphi_{i}(N_{i})}{N_{i}}}
\end{equation}
where $c_{ij}$ are complex numbers and $\varphi_{i}(N_{i})$ is an arbitrary
(complex) function with $\varphi_{i}(0)=0$, $\lim_{\epsilon \rightarrow 0}\frac
{\varphi_{i}(\epsilon)}{\epsilon}=1$, $|\varphi_{i}(1)|=1, \forall i \in S.$
We assume that $ |\varphi_{i}(N)| $ are bijective, monotonically increasing
functions  or that
$\varphi_{i}(N)=\frac {1-(-)^{N}}{2}$, implying $a_{i}^{2}=0$. The mappings (3)
generalize the mappings considered in \cite{j}\cite{fz}.\\
It is important to note that the number operators are preserved, i.e.
\begin{equation}
N_{i}^{(a)}=N_{i}^{(b)} \equiv N_{i}, \quad \forall i \in S.
\end{equation}
Then it is easy to find the corresponding deformed algebra:
\begin{equation}
\begin{array}{c}
a_{i}a_{j}=e^{c_{ji}-c_{ij}} a_{j}a_{i}  \quad i \neq j\\[4mm]
a_{i}a^{+}_{j}=e^{c_{ij}+c_{ji}^{*}}a^{+}_{j}a_{i},  \quad i \neq j \\[4mm]
a_{i}a^{+}_{i}=|\varphi_{i}(N_{i}+1)|  e^{\sum_{j}(c_{ij} +c_{ij}^{*})N_{j}}
e^{(c_{ii}+c_{ii}^{*})}
\\[4mm]
a^{+}_{i}a_{i}=|\varphi_{i}(N_{i})|  e^{\sum_{j}(c_{ij}+c_{ij}^{*})N_{j}}.
\end{array}
\end{equation}

Note that $a_{i}^{2}\neq 0$, unless
$b_{i}\varphi_{i}(N_{i})b_{i}\varphi_{i}(N_{i})=0
$ . For example,if $\varphi_{i}(N_{i})=\frac {1-(-)^{N_{i}}}{2}$, then
$a_{i}^{2}=0$,
implying the hard-core condition for the $i^{th}$ oscillator.
Generally , there are other mappings of Bose algebra, equations (1) and (2),
 but, in general, they do not have the number operators $ N_{i}^{(a)},$ and
equation (4) does not hold for other mappings than those in equation  (3).

We point out that the complete deformed algebra is associative owing
to the mapping of Bose algebra. The Fock space for the deformed
algebra is spanned by powers of the creation operators $ a_{i}^{+}$,$ i \in S,$
acting on the vacuum $|0>^{(a)}=|0>^{(b)} \equiv |0>.$ The states in the Fock
space are specified by the eigenvalues of the number operators $N_{i}$, namely
$|n_{1},n_{2},...n_{i}....>^{(a)}=|n_{1},n_{2},...n_{i}....>^{(b)}. $(If there
exists a number
$n_{i}^{(0)} \in {\bf N}$, such that $\varphi_{i}(n_{i}^{(0)})=0$, then
$N_{i}=0,1,...
(n_{i}^{(0)}-1)$).\\
States with unit norm are
\begin{equation}
\begin{array}{c}
|n_{1}.....n_{n}>=\frac{(a_{1}^{+})^{n_{1}}.......(a_{n}^{+})^{n_{n}}}{\sqrt
{[\tilde{
\varphi}_{1}(n_{1})]!.....[\tilde{\varphi}_{n}(n_{n})]! }}

e^{-\frac{1}{2}\sum_{j}\theta_{ji}(c_{ij}+c_{ij}^{*})n_{i}n_{j}}|0,0,....0>\\[6mm]
[\tilde{\varphi}(n)]! =\tilde{\varphi}(n)
\tilde{\varphi}(n-1)......\tilde{\varphi}(1)\\[4mm]
\tilde{\varphi}_{i}(n_{i})=|\varphi_{i}(n_{i})| e^{(c_{ii}+c_{ii}^{*})n_{i}}.
\end{array}
\end{equation}
where $\theta_{ij}$ is the step function. (For anyons in $(2+1)$ dimension
\cite{lsbdm}, $\theta $ is the angle function.)\\
Furthermore, the matrix elements of the operators $a_{i}, a_{i}^{+}$, $i \in
S$,
are
\begin{equation}
\begin{array}{c}
<....(n_{i}-1)....|a_{i}|....n_{i}...>=<...n_{i}....|a_{i}^{+}|....(n_{i}-1)....>^{*}\\[4mm]
=\sqrt{\varphi_{i}(n_{i})}e^{\frac{1}{2}\sum_{j}(c_{ij}+c_{ij}^{*})n_{j}}
\prod_{j\neq i}
\delta_{n_{j},n_{j}^{'}}.
\end{array}
\end{equation}
We also find for any $k=0,1,2,.... $ that
\begin{equation}
\begin{array}{c}
(a_{j}^{+})^{k}
(a_{j})^{k}=\frac{[\tilde{\varphi}_{j}(N_{j})]!}{[\tilde{\varphi}_{j}(N_{j}-k)]!}
e^{k\sum_{l\neq j}(c_{jl}+c_{jl}^{*})N_{l}}\\[4mm]
(a_{j})^{k}
(a_{j}^{+})^{k}=\frac{[\tilde{\varphi}_{j}(N_{j}+k)]!}{[\tilde{\varphi}_{j}(N_{j})]!}
 e^{k\sum_{l\neq j}(c_{jl}+c_{jl}^{*})N_{l}}
\end{array}
\end{equation}
The norms of arbitrary linear combinations of the states in equation (6)
in the Fock space corresponding to deformed algebra, are positive definite
owing to the mapping of Bose algebra, equations (1) and (2). Namely,
$|n_{1},n_{2},...n_{i}....>^{(a)}=\\
=|n_{1},n_{2},...n_{i}....>^{(b)} \equiv |n_{1},n_{2},...n_{i}....>$.

This class of deformed multimode  oscillator algebras comprises
multimode\\ Biedenharn-Macfarlaine \cite{bm}, Aric-Coon \cite{ac}, two-(p,q)
parameter \cite{cj}, Fermi,
generalized Green's \cite{mp}\cite{ok}, as well as anyonic \cite{lsbdm} and
Pusz-Woronowicz (with or
without the hard-core condition) oscillators  covariant under the $SU_{q}(n)$
$(SU_{q}(n|m))$ algebra (superalgebra) \cite{j}.

Non-isomorphic (non-equivalent) algebras are classified by different matrix
elements given by (7), i.e. with the functions $ g_{i}(n_{1},n_{2},...n_{n})=
|\varphi_{i}(n_{i})|e^{\sum_{j}(c_{ij}+c_{ij}^{*})n_{j}}.$ It is important to
mention that there are mappings of Bose algebra which do not preserve the
relation
$N_{i}^{(a)}=N_{i}^{(b)}$,  given by (4). Moreover, there are mappings of Bose
algebra
for which the number operators $N_{i}^{(a)} $do not even exist. Such an example
is the exchange algebra presented in \cite{bp}.

Finally, we construct and discuss the number operators $N_{i}^{(a)}$.Let
$\varphi_{i}(n_{i})$
be bijective mappings. Then, using equation (3) we have
\begin{equation}
\begin{array}{c}
b_{i}=a_{i} e^{-\sum_{j}c_{ij}N_{j}}\sqrt{\frac{N_{i}}{\varphi_{i}(n_{i})}},
\quad \forall i \in S\\[4mm]
N_{i}=b_{i}^{+}b_{i}.
\end{array}
\end{equation}
The spectra of $ N_{i}^{(a)} $ and $N_{i}^{(b)} $ coincide. In this case,
\begin{equation}
a_{i}^{+}a_{i}=g_{i}(N_{1},N_{2},...N_{i} ...)=|\tilde {\varphi}_{i}(N_{i})|
 e^{\sum_{j\neq i}(c_{ij}+c_{ij}^{*})N_{j}}.
\end{equation}
Let us denote $ a_{i}^{+}a_{i}=x_{i}$, and remark that $x_{i}$ commute with
any $N_{j}$ and among themselves. Then the $ N_{i} $ operators can be
written as
\begin{equation}
\begin{array}{c}
N_{i}=N_{i}(x_{1},...x_{n})=\sum_{k=0}^{\infty} \sum_{(i_{1}...i_{k}) }
\frac{1}{k!} (\frac{\partial^{k}N_{i}}{\partial x_{i_{1}}..\partial
x_{i_{k}}})_{x=0}
  (x_{i_{1}}x_{i_{2}}.....x_{i_{k}})\\[4mm]
N_{i}(0,0,..0)=0, \quad \forall i \in S.
\end{array}
\end{equation}
The coefficients in the Taylor expansion can be obtained  from (10)
\begin{equation}
g_{i}(N_{1},N_{2},...N_{i} ...)=x_{i}, \quad \forall i \in S
\end{equation}
namely from
\begin{equation}
(\frac{\partial^{k} g_{i}(N_{1},N_{2},...N_{i} ...)}{\partial
x_{j_{1}}..\partial x_{j_{k}}})_{x=0} =
\delta_{k1}\delta_{ji} .
\end{equation}
These equations give a set of recurrence relations for the coefficients in the
Taylor expansion of $N_{i}$ as a function of the variables $x_{j}$, $j \in S$.
For example,
\begin{equation}
\begin{array}{c}
c^{(i)}_{0}\equiv (N_{i})_{x=0}=0\\[4mm]
c^{(i)}_{j}\equiv (\frac{\partial N_{i}}{\partial
x_{j}})_{x=0}=\frac{1}{(\frac{ \partial g_{i}}{\partial N_{i}})_{0}}
\delta_{ij}\\[4mm]
c^{(i)}_{jk}\equiv \frac{1}{2}(\frac{\partial^{2} N_{i}}{\partial x_{j}
\partial x_{k}})_{x=0}=
 - \frac{1}{2} (\frac{\partial^{2} g_{i}}{\partial N_{j} \partial\ N_{k}})_{0}
   c^{(i)}_{i} c^{(j)}_{j} c^{(k)}_{k} .
\end{array}
\end{equation}
Specially
\begin{equation}
\begin{array}{c}
c^{(i)}_{j}=\frac{1}{(\frac{d \varphi_{i}}{d N_{i}})_{0}} \delta_{ij}
\\[4mm]
c^{(i)}_{ij}= -\frac{1}{2} (c_{ij}+c_{ji}^{*}) c^{(i)}_{i} c^{(j)}_{j}\\[4mm]
 c^{(i)}_{jk}=0,  \quad j\neq i, \quad k\neq i
\end{array}
\end{equation}
where  $g_{i}(N_{1},N_{2},...N_{i} ...)$ is given by equation (10).
We also mention the following result.

If there exists a continuous mapping defined by (3) from Bose oscillators to de
formed oscillators, then the number operators exist and can be written in
the form
 \begin{equation}
N_{i}=x_{i} [c^{(i)}_{i} + \sum_{j}  c^{(i)}_{j} x_{j} +...]
\end{equation}
where $x_{j}=a_{j}^{+}a_{j}$.

For example, let us consider the n-mode Pusz-Woronowicz
oscillator algebra\cite{pw} (of Bose type, covariant under the $SU_{q}(n)$
quantum algebra), $ q$ $\in $ {\bf R} :
\begin{equation}
\begin{array}{c}
a_{i} a_{j}= q^{sgn(j-i)} a_{j}a_{i}\\[4mm]
a_{i} a^{+}_{j}=q  a^{+}_{j} a_{i}, \quad  i \neq j \\[4mm]
a_{i}a^{+}_{i}=1+(q^{2}-1)\sum_{j}\theta_{ij}a^{+}_{j}a_{j}+q^{2}
a^{+}_{i}a_{i}.
\end{array}
\end{equation}
There exist mappings to the Bose algebra (1) and (2) with $c_{ij}=\theta_{ij}
\ln q$, \\$\varphi (N)=\frac{q^{2N}-1}{q^{2}-1}$ (see Table 1). Hence, we
can calculate the coefficients in the expansion (16). Using equation (15)
we obtain
\begin{equation}
\begin{array}{c}
c^{(i)}_{j}=\frac{q^{2}-1}{\ln q^{2}} \delta_{ij}\\[4mm]
c^{(i)}_{ij}= - \frac{1}{2} \frac{(q^{2}-1)^{2}}{\ln q^{2}} (\delta_{ij} +
\theta_{ij}).
\end{array}
\end{equation}
When $q^{2} \rightarrow 1$, then $ c^{(i)}_{i}\rightarrow 1$, whereas
$c^{(i)}_{ij} \rightarrow 0$, etc., and $N_{i}  \rightarrow a^{+}_{i}a_{i}$.
However, when $q^{2} \rightarrow 0$, all coefficients diverge and the
expansion (16) is not valid. Note that when $q=0$, the mapping (9)
becomes singular, but one can still define the corresponding number operators.
For $q=0$, the Pusz-Woronowicz algebra (17) reduces to
\begin{equation}
\begin{array}{c}
 a_{i} a_{j}=0,   \quad   i < j\\[4mm]
a_{i} a^{+}_{j}=0,  \quad i \neq j\\[4mm]
a_{i}a^{+}_{i}=1- \sum_{j}\theta_{ij}a^{+}_{j}a_{j}.
\end{array}
\end{equation}
Therefore we also present the number operators in another form that holds
for an arbitrary algebra having number operators, and holds even for
 $q=0 $\cite{mp}:
\begin{equation}
N_{i}=a^{+}_{i}a_{i} + \sum_{k=1}^{\infty} \sum_{\pi \in S_{k}}
\sum_{j_{1}...j_{k}}
d_{\pi (j_{1}...j_{k}),j_{1}...j_{k}}\\[2mm]
 a^{+}_{\pi (j_{k})}.......a^{+}_{\pi (j_{1})} a^{+}_{i}a_{i}
a_{j_{1}}....a_{j_{k}}
\end{equation}
where $S_{k}$ denotes the permutation group and $d_{\pi
(j_{1}...j_{k}),j_{1}...j_{k}}$
are the coefficients of the expansion.\\
For the Pusz-Woronowicz algebra the number operators in the above form (see
also \cite{j}) are given by
\begin{equation}
N_{i}=\sum_{k=0}^{\infty} \sum_{j=0}^{k} c_{k,j} \sum_{p_{1}....p_{j}}
\theta_{ip_{1}}.....\theta_{ip_{j}} a^{+}_{p_{j}}..... a^{+}_{p_{1}}
(a^{+}_{i})^{k-j+1}
(a_{i})^{k-j+1} a_{p_{1}}......a_{p_{j}}
\end{equation}
with the conditions $c_{0,0}=1$, $c_{0,-j}=0$, $j>0$, and with the recurrence
relations
\begin{equation}
c_{k+1,k+1-j}=\frac{(1-q^{2})(1-q^{2j})}{1-q^{2(j+1)}} c_{k,k+ 1-j} +
q^{2j}(1-q^{2}) c_{k,k-j}.
\end{equation}
Starting with $c_{0,0}=1$, we find that
\begin{equation}
\begin{array}{c}
c_{k,1}=c_{k,k}=(1-q^{2})^{k}\\[4mm]
c_{k,0}=\frac{(1-q^{2})^{k+1}}{1-q^{2(k+1)}}.
\end{array}
\end{equation}
In the limit $q^{2} \rightarrow 1$, the number operator is $N_{i}
=a^{+}_{i}a_{i}$.
In the limit $q=0$, all coefficients are $c_{k,j}=1$ , $\forall k,j$ .
This result is similar to that found by Greenberg \cite{g} for quons with
$q=0$. Finally, for $\varphi_{i}(N_{i})=\frac {1-(-)^{N_{i}}}{2}$, the
number operators, $ N_{i}^{(a)} \neq  N_{i}^{(b)}$, are simply
\begin{equation}
N_{i}^{(a)}=a^{+}_{i}a_{i}=b_{i}^{+} b_{i} [1-\theta (n_{i}-1)].
\end{equation}

We conclude that we unify and generalize the results for the states in the Fock
space
and the number operators for the mappings of multimode Bose algebra
presented in Table 1.
%
%
%***************************ACKNOWLEDGEMENTS****************************newpage
\newpage
{\bf Acknowledgments}

This work was supported by the joint Croatian-American contract NSF JF 999 and
the Scientific Fund of the Republic of Croatia.
%
%************************END OF ACKNOWLEDGEMENTS**************************
%
%
%*************************REFERENCES**************************************

%
%
%************************END OF REFERENCES*******************************
%
%
%****************************TABLE****************************************
%
\newpage

\begin{table}[h]
\caption{Parameters $c_{ij}$ and $\varphi_{i}(N_{i})$ for the various deformed
algebras.We use definition
$[x]_{p}=\frac{p^{x}-1}{p-1}.$}
\vskip1truecm
\begin{tabular}{| c | c |c |}  \hline

type of oscillators                                           & $c_{ij}$ &
$\varphi_{i}(N_{i})$ \\ \hline\hline

Bose                                            & $i\pi \lambda_{ij}$,
$\lambda_{ij}- \lambda_{ji} \in 2{\bf Z}$, $\forall i,j$ &   $N_{i}$ \\ \hline

Fermi                              &  $i \pi \lambda_{ij}$, $\lambda_{ij}-
\lambda_{ji} \in 2{\bf Z}+1$, $\forall i,j$  &  $[N_{i}]_{-1}$  \\ \hline

Green's oscillators \cite{mp}\cite{ok}  &    $i \pi \lambda_{ij}$,
$\lambda_{ij}- \lambda_{ji} \in {\bf Z} $, $\forall i,j$ & $[N_{i}]_{\pm 1}$
\\ \hline

Anyonic-type \cite{fz} & $i \pi \lambda_{ij}$, $\lambda_{ij}- \lambda_{ji} \in
{\bf R}$,  $\forall i,j$ &$[N_{i}]_{\pm 1}$  \\ \hline

Anyons  \cite{lsbdm} & $i \lambda \theta_{ij}$, $\lambda  \in {\bf R}$,
$\theta_{ij}$ is angle  & $[N_{i}]_{\pm 1}$  \\ \hline

Pusz-Woronowicz (Bose) \cite{pw}    & $  \theta_{ij} \ln q$, $q\in {\bf R}$  &
 $[N_{i}]_{q^{2}}$  \\ \hline

Pusz-Woronowicz (Fermi) \cite{pw}  & $  \theta_{ij} \ln (-q)$, $q\in {\bf R}$
&  $[N_{i}]_{-1}$  \\ \hline

$SU_{q}(n|m)$-covariant (Bose) oscillators  \cite{j} &  $\theta_{ij} \ln q$,
$q\in {\bf R}$, $ i \leq n$   &  $[N_{i}]_{q^{2}}$  \\ \hline

$SU_{q}(n|m)$-covariant (Fermi) oscillators \cite{j} &  $  \theta_{ij} \ln
(-q)$, $q\in {\bf R}$, $n+1\leq i $   &  $[N_{i}]_{-1}$  \\ \hline

Biedenharn-Macfarlane \cite{bm}     &  $-\frac{1}{2} \delta_{ij} \ln q_{i}$,
$q_{i}\in {\bf C}$  &  $[N_{i}]_{q_{i}^{2}}$  \\ \hline

Arik-Coon \cite{ac}                           &  0,  $q_{i}\in {\bf C}$ &
$[N_{i}]_{q_{i}}$ \\ \hline
\end{tabular}
\end{table}

%****************************END TABLE****************************************
%

%
\end{document}